\documentclass[twocolumn,showpacs,superscriptaddress,amssymb,10pt]{revtex4-2}

\usepackage{graphicx}
\usepackage{dcolumn}
\usepackage{bm}

\input{epsf}
\usepackage{epsfig}
\usepackage{array}
\usepackage{longtable}
\usepackage{rotating,booktabs}
\usepackage{CJK}
\usepackage[marginal]{footmisc}
\usepackage{booktabs}
\usepackage{multirow}
\usepackage{booktabs,threeparttable}
\usepackage{bm}
\usepackage{amsmath}
\usepackage{graphicx}
\usepackage{color}

\begin{document}


\title{Stepwise ionization of Mo$^{14+}$ ions in EBIT$:$ The importance of the metastable level} 


\author{Cunqiang Wu}
\affiliation{Key Laboratory of Atomic and Molecular Physics $\&$ Functional Materials of Gansu Province, College of Physics and Electronic Engineering, Northwest Normal University, Lanzhou 730070, P.R. China}

\author{Xiaobin Ding}
\email{dingxb@nwnu.edu.cn}
\affiliation{Key Laboratory of Atomic and Molecular Physics $\&$ Functional Materials of Gansu Province, College of Physics and Electronic Engineering, Northwest Normal University, Lanzhou 730070, P.R. China}
\affiliation{Gansu International Scientific and Technological Cooperation Base of Laser Plasma Spectroscopy, Lanzhou 730070,China}

\author{Qi Guo}
\affiliation{Shanghai EBIT laboratory, and Key Laboratory of Nuclear Physics and Ion-Beam Application (MOE), Institute of Modern Physics, Fudan University, Shanghai 200433, China}

\author{Ke Yao}
\affiliation{Shanghai EBIT laboratory, and Key Laboratory of Nuclear Physics and Ion-Beam Application (MOE), Institute of Modern Physics, Fudan University, Shanghai 200433, China}

\author{Jialin Liu}
\affiliation{Shanghai EBIT laboratory, and Key Laboratory of Nuclear Physics and Ion-Beam Application (MOE), Institute of Modern Physics, Fudan University, Shanghai 200433, China}

\author{Yunqing Fu}
\affiliation{Shanghai EBIT laboratory, and Key Laboratory of Nuclear Physics and Ion-Beam Application (MOE), Institute of Modern Physics, Fudan University,	Shanghai 200433, China}
\author{Chenzhong Dong}
\affiliation{Key Laboratory of Atomic and Molecular Physics $\&$ Functional Materials of Gansu Province, College of Physics and Electronic Engineering, Northwest Normal University, Lanzhou 730070, P.R. China}
\affiliation{Gansu International Scientific and Technological Cooperation Base of Laser Plasma Spectroscopy, Lanzhou 730070,China}

\date{\today}

\begin{abstract}
The visible spectrum of Mo$^{15+}$ ions was measured using a high-temperature superconducting electron-beam ion trap at the Shanghai EBIT Laboratory, with an electron beam energy $E_{e}$=400eV, significantly lower than the ionization potential (IP=544.0 eV) of Mo$^{14+}$ ions in ground state. To expound on the experiment, the energy level structure, radiative transition properties, electron-impact excitation, and electron-impact ionization cross section for both the ground state and low-lying excited state of the Mo$^{14+}$ ions were calculated using Dirac–Fock-Slater method with a local central potential and distorted wave approximation. The results demonstrated reasonable agreement with both available experimental and theoretical data. Through an analysis of the related atomic processes of Mo$^{14+}$ ion, a scenario involving the stepwise ionization of the metastable state 3p$^{6}$3d$^{9}$4s was proposed to explain the presence of the Mo$^{15+}$ ions with a lower energy of incident electron. Finally, the significance of the metastable levels in the ionization of Mo$^{14+}$ ions is highlighted. 
\end{abstract}


\maketitle

\section{\label{Sec:Introduction}Introduction}

The study of the metastable level structures and properties of highly charged ions (HCIs) hold importance, not only in testing atomic physics theory that describes the interaction of atoms or ions with multipole radiative fields \cite{Ralchenko2011,Guise2014,Saito2015,Mueller2018,Sakaue2019}, but also the radiative emissions from these metastable levels prove crucial for the effective application of density diagnostics in astrophysical and laboratory plasmas \cite{Moehs2000,Steinbruegge2022,Chen_2023}. The metastable levels of atoms or ions are characterized by an inability to decay to lower energy levels via electric dipole or Auger processes, constrained by the limitations of selection rules, thereby extending the lifetime of the metastable level to millisecond or more \cite{Kim2002,Brenner2007,Brandt2022}. Certain long-lived metastable levels, which create an unperturbed, well-defined, ideal two-level system, hold potential applications in quantum information, plasma-assisted laser spectroscopy \cite{Schuessler2020}, penning-trap mass  spectrometry \cite{Kimura2023a}, and investigations into the variation of fundamental constants \cite{Safronova2014,Safronova2014a,Safronova2014b}. Furthermore, long-lived metastable states with high order forbidden transitions serve as valuable candidates for studying hyperfine-induced transitions and magnetically-induced transitions,  as new decay channels can significantly influence their lifetimes \cite{Schippers2007,Traebert2007}.

The long-lived metastable state can be considered analogous to the ground state, thereby possessing similar properties and behaviors to the ground state. In hot plasma environments,  Novel electron-impact channels for excitation, deexcitation, ionization, and recombination may open for the metastable level \cite{Kobayashi2014}. In magnetic confinement fusion devices like the International Thermonuclear Experimental Reactor (ITER), Tungsten (W) and Molybdenum (Mo) serve as the divertor and plasma facing component materials. Research has shown that the population of metastable states has a significant affects measurements of tungsten erosion from plasma facing components and impurity transport \cite{Quinet_2010,Johnson_2020}. Studies from the electron beam ion trap (EBIT) further underscore that the presence of a long-lived metastable state with reduced ionization energy can instigate a stepwise ionization process. This process has a significant impact on establishment and progression of plasma's charge state equilibrium, as well as the characteristics of its emission spectrum \cite{Li2015,Lu2019}.

The detection and manipulation of the charge state of HCIs are imperative for a range of plasma applications. An EBIT \cite{Fu2010,Nakamura2008}, chiefly comprising quasi-monochromatic electron beams, serves as a powerful tool for investigating the level structure, spectroscopic properties, and atomic processes in plasmas with metastable state of HCIs \cite{Sakoda2011,Torretti2017}. Recent experimental and theoretical \cite{Lu2019,Kimura2020,Lu2021,Yan2022,Kimura2023} research has indicated the undeniable significance of diverse atomic processes induced by the metastable states of HCIs.  Investigations into several HCIs reveal that ionization processes initiated by metastable states can be observed by monitoring the emission spectra, which showcases incident electron energies significantly below the threshold required for  producing HCIs from the ground state of the preceding low charge state ions \cite{Sakoda2011,Qiu2014,Windberger2016,Torretti2017,Lu2019,Kimura2020,Lu2021,Yan2022,Kimura2023}. Numerous ions possessing metastable states have the potential to prompt complex ionization balance by stepwise ionization pathways. While ionization of metastable states has been investigated, such instances remain infrequent, thereby indicating a necessity for a more systematic and quantitative analysis of stepwise ionization in HCIs.

In this work, the Mo$^{15+}$/Mo$^{14+}$ ratio, the charge state distribution of molybdenum ions, and the visible spectrum of Mo$^{15+}$ ion are measured at the Shanghai low-energy EBIT, CUbic-eBIT (CUBIT) \cite{Xiao2012,He2022}. The observed visible spectrum is identified as the M1 transition occurring between the doublet levels of the ground configuration of Mo$^{15+}$ ion. The ionization energy of Mo$^{14+}$ ions stands at 544.0 eV \cite{NIST_ASD}, and the experimental results demonstrate that the M1 transition peak emerges where lower electron beam energy is deployed. Consequently, by utilize the Dirac-Fock-Slaterrmcdhf (DFS) method with a local central potential in the Flexible Atomic Code (FAC) \cite{Gu2008,Gu2004}, we compute the levels structure, radiative transition probability, and cross sections for electron collisions of Mo$^{15+}$ and Mo$^{14+}$ ions. Additionally, the GRASP2K code \cite{Joensson2013,FroeseFischer2019}, underpinned by the multi-configuration Dirac-Hartree-Fock (MCDHF) theory, is also utilized to compute the transition wavelength. Ultimately, a viable collisional-radiative model (CRM) was assembled and resolved, taking the the metastable levels into account. Furthermore, we postulated that the production of the Mo$^{15+}$ ion could be attribute to the ionization of metastable Mo$^{14+}$ ions. 

\section{\label{Sec:EXPERIMENTAL SETUP}Experimental setup}

\begin{figure*}[htbp]
	\centering
	\includegraphics[height=8.0cm,width=14cm]{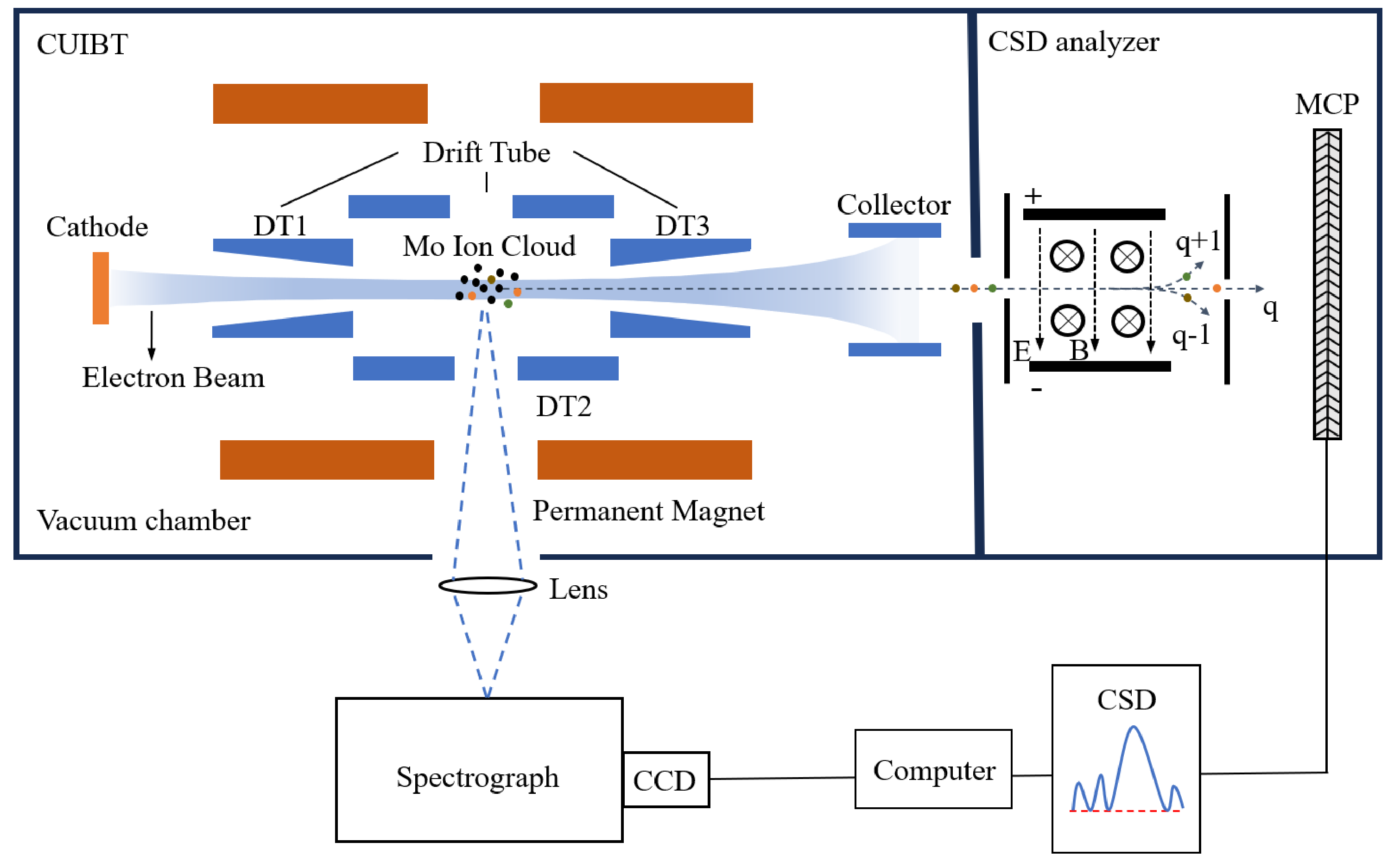}
	\setlength{\abovecaptionskip}{-0cm}
	\caption{\label{Fig-1} Scheme of the experimental setup.}
\end{figure*}

A schematic of the experimental setup is depicted in Fig.\ref{Fig-1}. The setup primarily consists of the CUBIT \cite{He2022,Li2022,Wang2022}, a spectrograph, a charge-coupled device (CCD) camera, and a charge state distribution (CSD) analyzer. CUIBT is a room temperature permanent magnet electron beam ion trap where highly charged ions are produced through electron impact ionization. Inside CUIBT, an electron beam emitted from a LaB$_{6}$ cathode is compressed by a 0.56T magnetic field provided by a permanent magnet as it passes through the center drift tube. The combination of the magnetic field and the electron beam space charge provides radial trapping, while the voltage applied on the drift tubes forms the axial electrostatic trapping potential. The electron beam energy is controlled by adjusting the potential difference between the cathode and the central drift tube, enabling the production of desired charged states of ions.

As a light source, visible spectra emitted from CUBIT can be observed with a CCD camera mounted on a Czerny-Turner type visible spectrometer. Additionally, ions in the trap region can be extracted in a pulse mode by lowering potential on the end drift tube. These extracted ions are then transported to the CSD analyzer, whrer ions with different charge to mass ratios are separated using orthogonal electric and magnetic fields. Only ions with specific charge to mass ratio can be detected by a microchannel plate (MCP). The CSD of ions is measured by scanning the voltage applied on the analyzer.

 The spectra of molybdenum ions measured in experiment are shown in Fig.\ref{Fig-2}(a). Simultaneously, the corresponding charge state distributions (CSDs) of molybdenum ions were measured in a pulsed mode. In this mode, the operation of the power system was regulated by the clock signal. When the end drift tube was at high potential for about 1.6s, HCIs were created and trapped in the middle drift tube, and the light emitted by the trapped ions was observed during this period. Subsequently, the voltage on the end drift tube was rapidly lowered to a low potential to open the trap. Therefore, the molybdenum ions were promptly dumped and extracted, and a pulsed HCIs bunch was detected. Fig.\ref{Fig-2}(b) shows the measured CSDs at the same electron energy.

\section{\label{Sec:Theory-and-computation}Theory and computation}

All pertinent atomic data, including energy levels, cross sections for electron-impact excitation/ionization and radiative recombination, alongside radiative transition probabilities, have been calculated utilizing FAC \cite{Gu2008}. The frequency-dependent Breit interaction in the zero-energy limit for the exchanged photon and higher-order QED effects such as self-energy and vacuum polarization corrections (treated in the screened hydrogenic approximation) were added in a subsequent relativistic configuration interaction calculation \cite{Koziol2020}. The collisional-radiative model implemented in FAC has been used to analyze the spectra and to predict the charge-state distributions \cite{Gu2008}.  

The current CRM encompasses 450 and 212 excited levels, originating from 3p$^{6}$3d$^{10}$, 3p$^{6}$3d$^{9}4l$ ($l$=0--3), 3p$^{6}$3d$^{8}nl^{2}$, 3p$^{4}$3d$^{10}nl^{2}$ ($n$=4--6, $l$=0--3), 3p$^{5}$3d$^{9}4l5l$ ($l$=0--3), 3p$^{5}$3d$^{9}4l^{2}$ ($l$=0--3), 3s$^{1}$3p$^{5}$3d$^{10}4l^{2}$ ($l$=0--3), and 3p$^{6}$3d$^{9}$, 3p$^{5}$3d$^{10}$, 3p$^{6}$3d$^{8}nl$, 3p$^{6}$3d$^{7}nl^{2}$ ($n$=4--6, $l$=0--3), 3p$^{5}$3d$^{9}4l$ ($l$=0--3) 3p$^{5}$3d$^{9}nl$ ($n$=4,5, $l$=0--1), 3p$^{5}$3d$^{8}4l^{2}$ ($l$=0--3), 3p$^{6}$3d$^{7}4l5l'$ ($l$=0--1, $l'$=1--4 ) configurations for Mo$^{14+}$ and Mo$^{15+}$, correspondingly. The populations of the excited levels are derived by solving the rate equations for all excited states. In the current model, we postulate the various levels of the ions are interconnected through only collisional and radiative process within the plasma. Within the context of low-density EBIT plasma, our primary considerations have been electron-impact excitation and deexcitation, electron-impact ionization, radiative transition, and radiative recombination. On the other hand, the other processes such as dielectronic recombination, three-body recombination, charge exchange, and ion escape can be disregarded. The following rate equation (\ref{Eq.1}) can be employed to describe the differential rate of the population of each excited level:

\begin{widetext}
	\begin{equation}
		\begin{aligned}
			\frac{d N_i}{d t}=\sum_{\substack{
					i \neq j}} Q_{i j} N_i N_e+\sum_{i>j} A_{i j} N_i+N_{+} N_e R_{+j}-\sum_{\substack{	i \neq j}} Q_{j i} N_j N_e \quad-\sum_{i<j} A_{j i} N_j-N_j N_e S_{j+} \label{Eq.1}.
		\end{aligned}
	\end{equation}
\end{widetext}

where, \textit{N$_{e}$} and \textit{N$_{i}$} represent the electron density and the populations of the \textit{i}-th level. \textit{$A$} is the radiative transition probabilities (including E1, E2, E3, M1, M2, and M3). \textit{$Q$}, \textit{$S$}, and \textit{$R$} denote the electron-impact excitation or deexcitation, electron-impact ionization, and radiative recombination rate coefficients, respectively. These rate coefficients can be derived by  by convoluting the corresponding cross sections on the free electron energy distribution function, described by the $\delta$ function for the monoenergetic electron beam of EBIT. The cross section are computed using the distorted wave approximation, while the inverse process are inferred in accordance with the detailed balance principle. Given the quasi-steady-state approximation $\frac{d N_{i}}{dt}=0$ and normalization condition $\sum_i N_i=1 $, the population of each level can be ascertained within the specific electron density and temperature.

\section{\label{Sec:Results-and-discussion}Results and discussion}

\begin{figure}[ht]
	\centering
	\includegraphics [height=6.8cm,width=8.5cm] {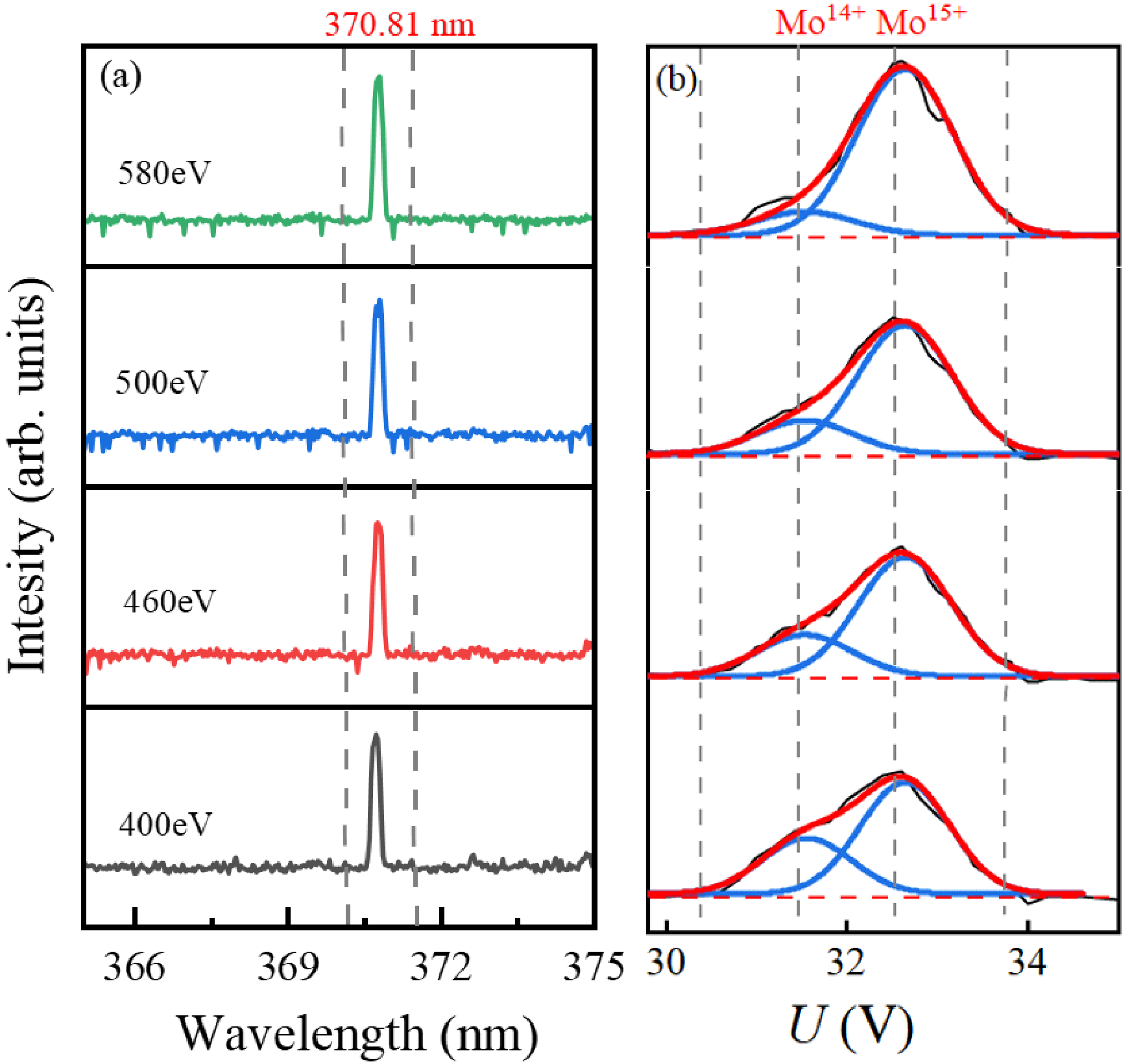}
	\caption{\label{Fig-2}(a) Spectra of molybdenum ions around 370 nm recorded at electron-beam energies of 400, 460, 500, and 580 eV. The line at 370.81 nm is the 3p$^{6}$3d$^{9}$ ($^{2}$D$_{3/2}$$\rightarrow$$^{2}$D$_{5/2}$) M1 transition of the Mo$^{15+}$ ion. (b) CSDs of Mo ions for the same electron energies. Black lines are the results of measurements, the solid red and blue lines are the Gaussian fits used to infer the CSDs, and the dashed red lines show the baselines of the fits. Dashed purple lines indicates the centroid of the specific charged state.}
\end{figure}

Fig.\ref{Fig-2}(a) depicts the experimental spectra of molybdenum ions observed at electron-beam energies of 400, 460, 500and 580 eV, respectively. It is evident from Figure \ref{Fig-2}(a) that an isolated transition peak of the Mo$^{15+}$ ion is observed at a beam energy of 400eV with the wavelength of 370.81nm. Furthermore, Fig.\ref{Fig-2}(b) illustrates that the abundance of Mo$^{15 +}$ predominates during this period. Utilizing the MCDHF and DFS methods, we estimate the transition wavelength and compare it with extant experimental \cite{NIST_ASD} and theoretical \cite{Guo2016,Nandy2021} results, as presented in Table \ref{Table1}. As illustrated in the table, the wavelength of the observed transition peak principally corresponds to the M1 transition wavelength of the ground state 3p$^{6}$3d$^{9}$ ($^{2}$D$_{5/2}$$\rightarrow$$^{2}$D$_{3/2}$) of the Mo$^{15+}$ ion. Notwithstanding, while the ionization energy of Mo$^{14+}$ ions is approximately 544.0 eV \cite{NIST_ASD}, experimental observation have depicted that Mo$^{15+}$ ions exist at significantly lower ionization energy, for instance when $E_{e}$= 400eV. In order to elucidate this phenomenon, a hypothetical framework of stepwise ionization of the metastable state 3p$^{6}$$3d^{9} 4s$ has been proposed to account for the presence of Mo$^{15+}$ ions with lower incident electron energy.

\begin{table}[htbp]
	\renewcommand\arraystretch{1.20}
	\setlength{\tabcolsep}{0.1cm}
	\caption{\label{Table1} Wavelength of the 3p$^{6}$3d$^{9}$ $^{2}$D$_{5/2}$$\rightarrow$$^{2}$D$_{3/2}$ transition in Co-like Mo.}
	\begin{ruledtabular}
		\begin{tabular}{cccccc}
			\multicolumn{2}{c}{This work (nm)} &  & \multicolumn{2}{c}{Other work (nm)} \\
			\cmidrule(lr){1-2} 
			\cmidrule(lr){4-5} 
			MCDHF & DFS & NIST\textsuperscript{\cite{NIST_ASD}} & RMBPT\textsuperscript{\cite{Guo2016}}  & RCC\textsuperscript{\cite{Nandy2021}} \\
			\hline
			371.85 & 372.87 & 370.81 & 371.29 & 371.30  \\
		\end{tabular}
	\end{ruledtabular}
\end{table}

\begin{figure}[ht]
	\includegraphics [height=7cm,width=9cm] {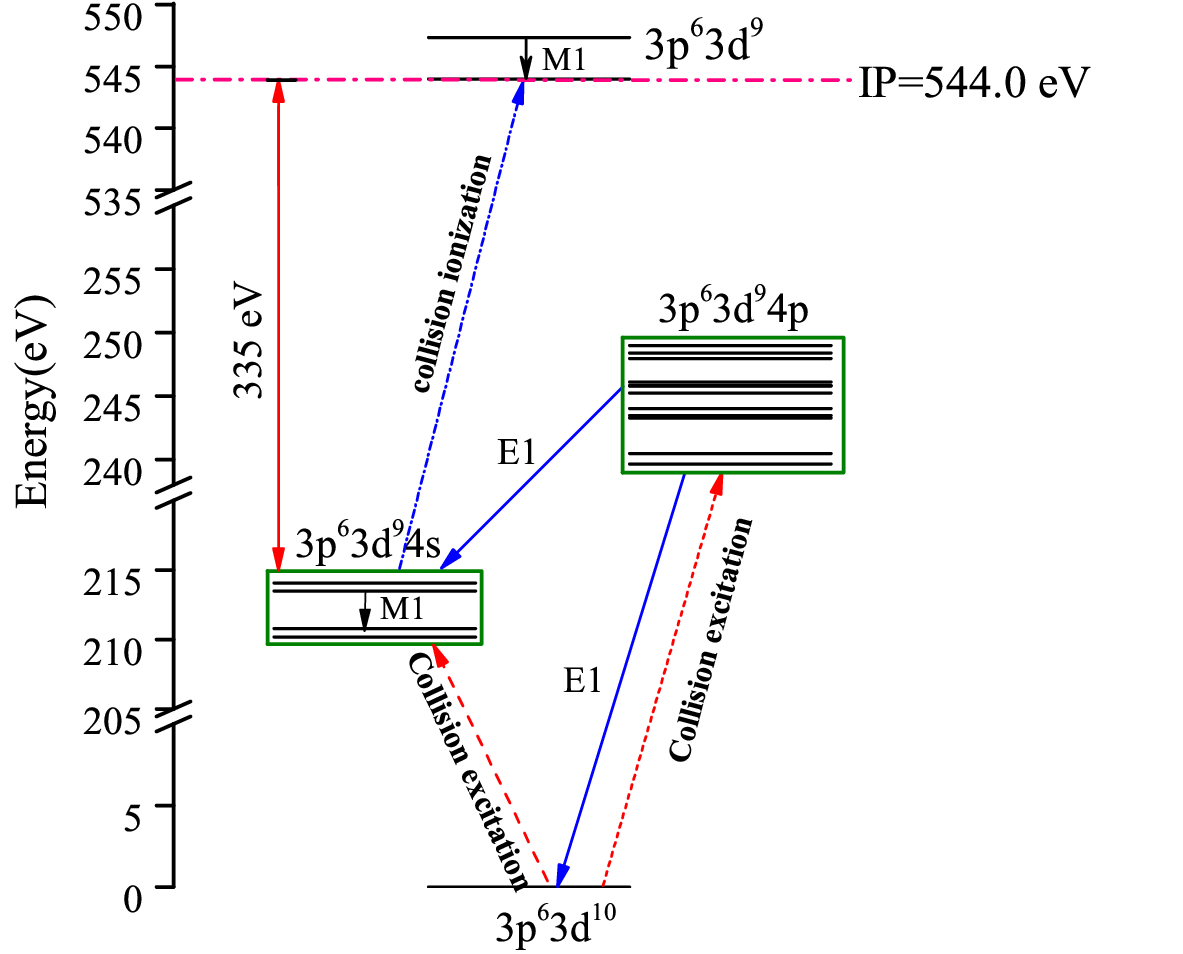}
	\setlength{\abovecaptionskip}{-0.2cm}
	\caption{\label{Fig-3}Energy level structure of Mo$^{14+}$ ion 3p$^{6}$3d$^{10}$, 3p$^{6}$3d$^{9}$4s, 3p$^{6}$3d$^{9}$4p and Mo$^{15+}$ ion 3p$^{6}$3d$^{9}$. The dash line represents electron-impact excitation, and the solid line represents radiation decay, the shot dot dash line represents collision ionization.}
\end{figure}

\begin{table*}[htbp]
	\renewcommand\arraystretch{1.20}
	\setlength{\tabcolsep}{0.02cm}
	\caption{\label{Table2} The excitation energy (in:eV) of the metastable state 3p$^{6}$3d$^{9}$4s of Mo$^{14+}$ ion and the ionization energy of ground state 3p$^{6}$3d$^{9}$ of Mo$^{15+}$ ion. Error=$\frac{(DFS-NIST)}{NIST}\times100\%$ }
	\begin{ruledtabular}
		\begin{tabular}{ccccccccc}
			&      &  &Present&    &Reference   &     &  \\
			\hline
			Charge state&Config & Level & DFS &NIST\textsuperscript{\cite{NIST_ASD}} & MCDF\textsuperscript{\cite{Brage2006}} & RPTMP\textsuperscript{\cite{Ivanova2021b}} & Error \\
			\hline
		Mo$^{14+}$&	 3p$^{6}$3d$^{10}$  &$ ^{1}$S$_{0}$   & 0     & 0      & 0      & 0      &     \\
			&3p$ ^{6}$3d$^{9}$4s &$^{3}$D$_{3}$   & 210.16 & 210.14 & 210.14 & 210.15 & 0.01\% \\
			&&$^{3}$D$_{2}$   & 210.78 &210.76  & 210.75 & 210.76 & 0.01\% \\
			&&$^{3}$D$_{1}$   & 213.46 & 213.47 & 213.47 & 213.47 & 0.01\% \\
		&	&$^{1}$D$_{2}$   & 214.04 & 214.05 & 214.04 & 214.05 & 0.01\% \\
			Mo$^{15+}$& 3p$^{6}$3d$^{9}$ &$^{2}$D$_{5/2}$ & 543.88 & 544.00 &        &        & 0.02\% \\
			&&$^{2}$D$_{3/2}$ & 547.20 & 547.34 &        &        & 0.03\% \\
		\end{tabular}
	\end{ruledtabular}
\end{table*}

Fig.\ref{Fig-3} illustrates the possible pathway for the production of Mo$^{15+}$ ions from Mo$^{14+}$ ions. In Fig.\ref{Fig-3}, the dash-dot line symbolizes the electron-impact excitation process, while the short dash-dot line corresponds to the electron-impact ionization process, and the solid line stands for the 3p$^{6}$3d$^{9}4p$ radiative decay process. As observed in Fig.\ref{Fig-3}, the the ionization energy required to produce  Mo$^{15+}$ from the metastable state 3p$^{6}$3d$^{9}4s$ is approximately 338 eV, which is less than the incident electron beam energy corresponding to the Mo$^{15+}$ ion spectral line used in the experiment, rendering this process of energetically feasible. Concurrently, Table \ref{Table2} displays the excitation energy for the low-lying excited configuration 3p$^{6}$3d$^{9}$4s of the Mo$^{14+}$ ion and the ionization energy of ground state configuration 3p$^{6}$3d$^{9}$ of Mo$^{15+}$ ion. The current results, obtained using the DFS methods, align remarkably well with NIST data \cite{NIST_ASD} and other theoretical results \cite{Brage2006,Ivanova2021b}. The maximum relative error of 0.03\% indicates the inclusion of significant configuration interactions in these calculations.

\begin{figure}[ht]
	\includegraphics [width=8.0cm] {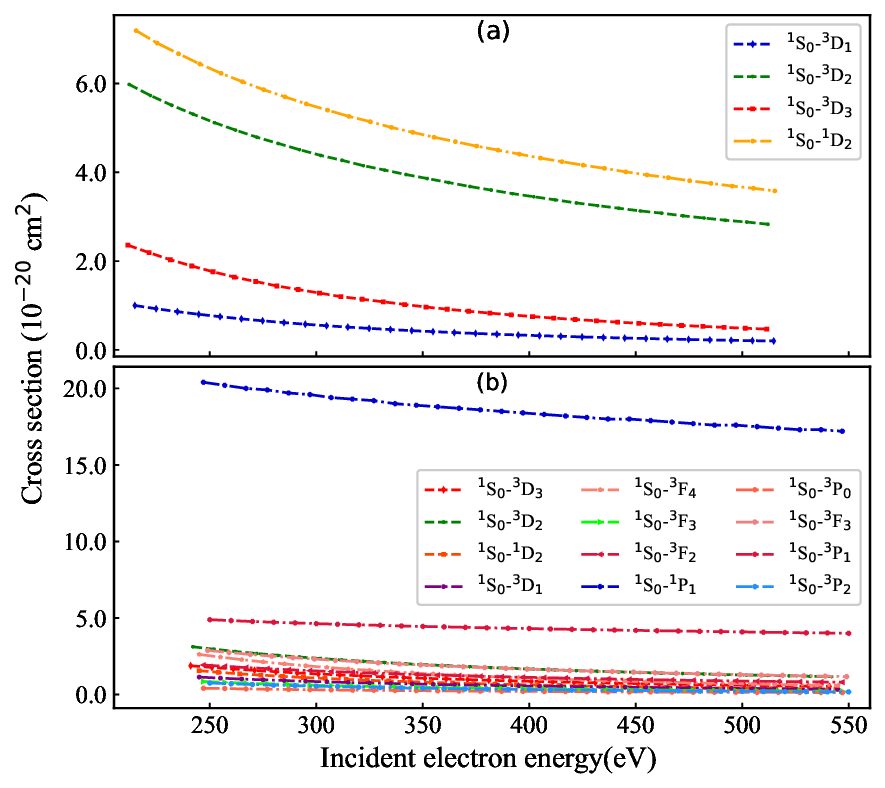}
	\setlength{\abovecaptionskip}{-0.1cm}
	\caption{\label{Fig-4}The electron-impact excitation cross section from the ground state 3p$^{6}$3d$^{10}$ to the low-lying excited state 3p$^{6}$3d$^{9}$4s (a) and the metastable state 3p$^{6}$3d$^{9}$4p (b) of Mo$^{14+}$ ions varies with electron energy.}
\end{figure}

Fig \ref{Fig-4} shows the electron-impact excitation cross section of the Mo$^{14+}$ ion 3p$^{6}$3d$^{10}$--3p$^{6}$3d$^{9}$4s (a) and 3p$^{6}$3d$^{10}$--3p$^{6}$3d$^{9}$4p (b) as a function of incident electron energy. Both the electron-impact excitation cross section from the ground state 3p$^{6}$3d$^{10}$ to the metastable state 3p$^{6}$3d$^{9}$4s and low-lying excited state 3p$^{6}$3d$^{9}$4p decrease monotonically with the increase in electron energy. However, Fig.\ref{Fig-4}(b) reveals that the cross section from the ground state 3p$^{6}$3d$^{10}$ ($^{1}$S$_{0})$ to the low-lying excited state 3p$^{6}$3d$^{9}$4p ($^{1}$P$_{1}$) is approximately three times larger than the cross section to 3p$^{6}$3d$^{9}$4s, while the rest of the cross sections are comparatively small. This observation suggests that the likelihood of $^{1}$S$_{0}$ being excited to $^{1}$P$_{1}$ by the electron-impact exceeds that of other states.

\vspace{-0.2cm}

\begin{figure}[ht]	
	\setlength{\abovecaptionskip}{-1.0cm}
	\includegraphics [width=8.5cm] {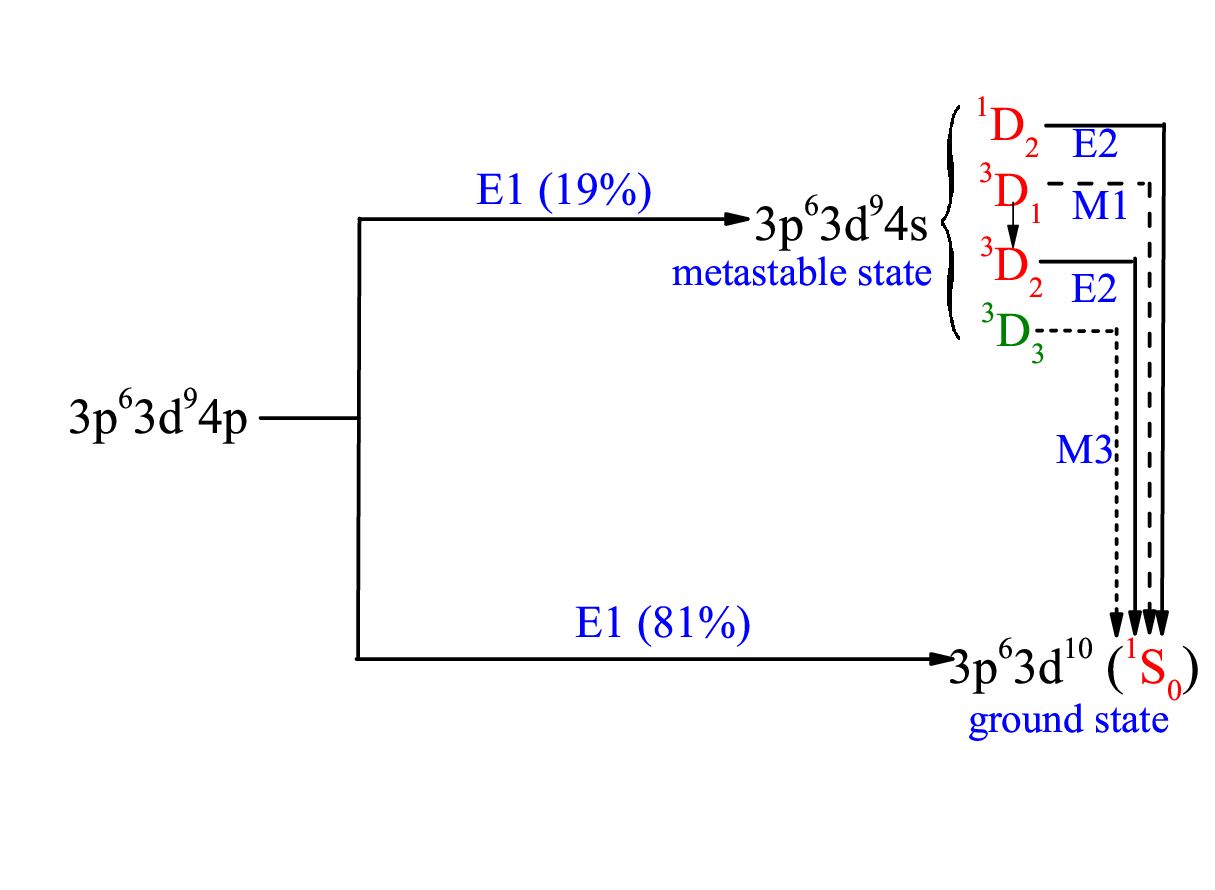}
	\caption{\label{Fig-5} Decay channels from low-lying excited state 3p$^{6}$3d$^{9}$4p of the Mo$^{14+}$ ions to the ground state 3p$^{6}$3d$^{10}$ and the metastable state 3p$^{6}$3d$^{9}$4s.}
\end{figure}

However, the low-lying excited state 3p$^{6}$3d$^{9}$4p will decay via E1 transition both to the ground state 3p$^{6}$3d$^{10}$ ($^{1}$S$_{0}$) and the metastable state 3p$^{6}$3d$^{9}$4s ($^{3}$D$_{3,2,1}$,$^{1}$D$_{2}$). The decay channels of the Mo$^{14+}$ ion's low-lying excited state and metastable state are portrayed in Fig.\ref{Fig-5}. In addition to the E1 transition to the ground state 3p$^{6}$3d$^{10}$, the low-lying excited state 3p$^{6}$3d$^{9}$4p also exhibits E1 decay to the metastable state 3p$^{6}$3d$^{9}$4s. The metastable states, including $^{3}D_{2}$ and $^{1}D_{2}$ exhibits E2 decay to the ground state, while $^{3}D_{1}$ demonstrate M1 decay to both ground state and internal energy levels $^{3}$$D_{2}$. However, $^{3}$D$_{3}$ shows only a weak M3 transition, which results in a substantial population of $^{3}D_{3}$.

\begin{figure}[ht]
	\setlength{\abovecaptionskip}{-0.1cm}
	\includegraphics [width=8.3cm] {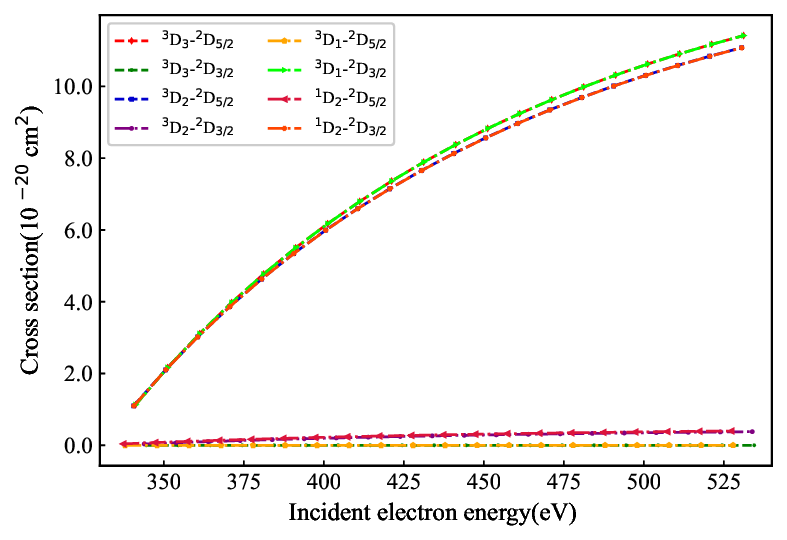}
	\caption{\label{Fig-6}The electron-impact ionization cross section of the Mo$^{14+}$ ion metastable state 3p$^{6}$3d$^{9}$4s varies with incident electron energy.}
\end{figure}

Fig.\ref{Fig-6} represents the electron-impact ionization cross section of the metastable state $3p^{6}3d^{9}4s$ of Mo$^{14+}$ ion. As the electron beam energy increases, the electron-impact ionization cross section gradually increases. Moreover, at the same electron energy, the $^{3}$D$_{3}$/$^{3}$D$_{2}$--$^{2}$D$_{5/2}$ and $^{3}$D$_{1}$/$^{1}$D$_{2}$--$^{2}$D$_{3/2}$ electron impact ionization cross section are an order of magnitude larger than other cross sections, and $^{3}$D$_{3}$--$^{2}$D$_{5/2}$ electron-impact ionization cross section is the largest. This demonstrates the increased likelihood of Mo$^{15 +}$ ions progressing through this channel.

\begin{table*}[htbp]
\renewcommand\arraystretch{1.2}
\setlength{\tabcolsep}{0.05cm}
\centering
\caption{\label{Table3}The lifetime and population of the Mo$^{14+}$ ion metastable 3p$^{6}$3d$^{9}$4s. The computation is conducted at electron-beam energy 400 eV and density $1.0\times10^{12}$/cm$^{3}$.}
\begin{ruledtabular}
	\begin{tabular}{ccccccccc}
		Upper&         Lower&          $\lambda$ (nm)   &Referce    &Type     & A$_{ij}(s^{-1})$  & Population$(\%)$  & Lifetime(ms)       \\ \hline
		$^{1}S_{0}$    &$-$                 &$-$        &$-$         &$-$      &$-$              & 96.24            & $-$              \\
		$^{3}D_{2}$    &$^{1}S_{0}$       &5.85         &5.88\textsuperscript{\cite{NIST_ASD}} &E2       &9.60E+06       & 0.005         &1.04E$-$04  \\
		$^{3}D_{1}$     &$^{1}S_{0}$       &5.78          & 5.83\textsuperscript{\cite{Safronova2006}} &M1           &4.88E$-$01           &0.45                  &4.13E+00                 \\
		&$^{3}D_{2}$       &       &       &M1  &2.42E+02           &                &                    \\
		$^{1}D_{2}$    &$^{3}D_{3}$       &        &     &M1    &3.22E+02           & 0.001               &7.30E$-$05       \\
		&$^{3}D_{2}$       &        &       &M1 &2.00E+01           &                    &                \\
		&$^{1}S_{0}$       &5.77          &5.79\textsuperscript{\cite{NIST_ASD}}    &E2   &1.37E+07           &                    &                \\
		$^{3}D_{3}$    &$^{1}S_{0}$       &5.87          &5.92\textsuperscript{\cite{Safronova2006}}    &M3        &3.83E$-$01           & 3.30               &2.85E+03        \\
	\end{tabular}	
\end{ruledtabular}
\end{table*}

Table \ref{Table3} presents the transition wavelength, transition probability, lifetime, and population of the Mo$^{14+}$ ion's metastable state 3p$^{6}$3d$^{9}$4s. The populations of each metastable levels are estimated by the collisional-radiative model, using an electron-beam energy 400 eV and density $1.0\times10^{12}$/cm$^{3}$. As observed in Table \ref{Table3}, $^{3}$D$_{3}$ possesses both a high population and a longer lifetime, and also exhibits the greatest electron-impact ionization cross section. The large cross section and high population increase the likelihood that ions will reach to the Mo$^{15+}$ ion's ground state through this route.

Based on the foregoing analysis, the presence of Mo$^{15+}$ ions, which were experimentally observed at a lower incident electron energy, can be attributed to the process where the ground state of Mo$^{14+}$ ion is collisionally excited to the metastable state 3p$^{6}$3d$^{9}$4s and low-lying excited state 3p$^{6}$3d$^{9}$4p by an electron. Subsequently, the low-lying excited state undergoes E1 decay to the metastable state. Furthermore, the metastable state is then collisionally ionized by an electron, resulting in the production of Mo$^{15+}$ ions.

\begin{figure}[ht]
	\setlength{\abovecaptionskip}{-0.1cm}
	\includegraphics [width=8.2cm] {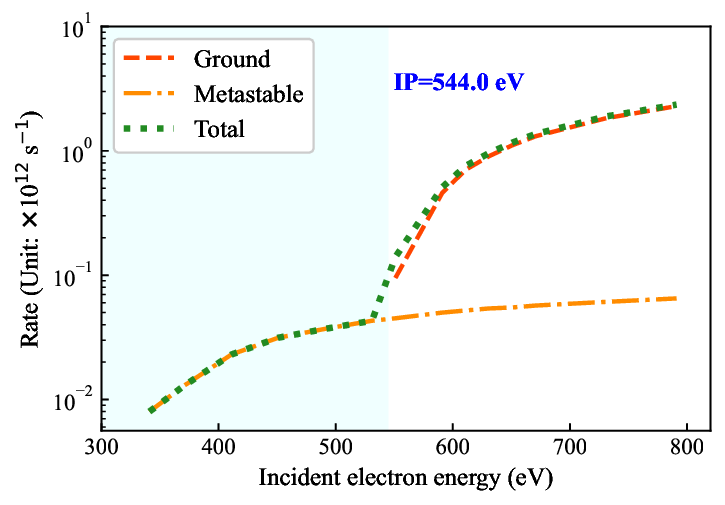}
	\caption{\label{Fig-7} The electron-impact ionization rate and total ionization rate of the Mo$^{14+}$ ion in both the ground 3p$^{6}$3d$^{10}$ ($^{1}$S$_{0}$) state and the metastable state 3p$^{6}$3d$^{9}$4s ($^{3}$D$_{3}$).}
\end{figure}

Fig.\ref{Fig-7} illustrates the electron-impact ionization rate for both the metastable state and ground state, as well as the total ionization rate of Mo$^{14+}$ ions. Below the ionization threshold of Mo$^{14+}$ ions (IP=554.0 eV), it is solely the stepwise ionization process of the metastable state that contributes to the total ionization rate. When over the ionization threshold, both the stepwise ionization of the metastable state and the direct ionization of the ground state simultaneously contribute to the total rate. The total ionization rate is determined by weighted summation of the electron-impact ionization rate for the metastable $(3.30\%)$ state and the ground state $(96.24\%)$.

\begin{figure}[ht]
	\includegraphics [width=8.0cm] {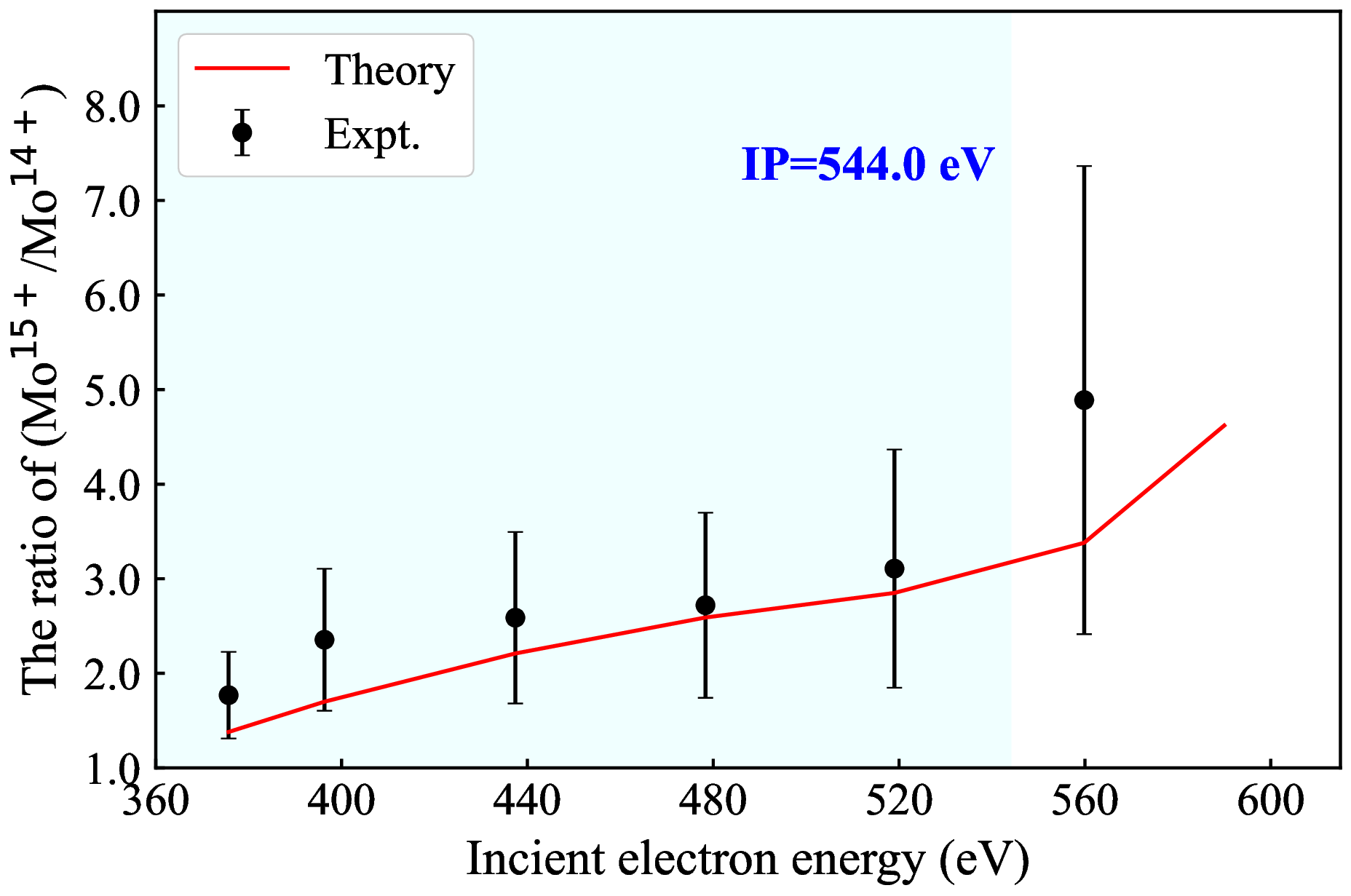}
	\caption{\label{Fig-8} Comparison of Mo$^{15+}$/Mo$^{14+}$ ion ratio between theoretical calculation and EBIT experimental measurement.}
\end{figure}

The experiment also observed the charge state ratio of Mo$^{15+}$ and Mo$^{14+}$ ions. When the incident electron energy is significantly lower than the ionization energy of Mo$^{14+}$ ion, the ratio between the two ions exceeds 1, a phenomenon primarily attributed to metastable stepwise ionization. Upon considering the various atomic processes associated with the aforementioned excitation and ionization channels, a suitable CRM was assembled to analyze the dependency of the ratio of these two ions on the incident electron energy. Good agreement was obtained between the experimental and the theoretical results, as portrayed in Fig.\ref{Fig-8}. Moreover, it is observed that when the incident electron energy surpasses the ionization energy, the ratio between the two ions experiences a sudden increase. This can be primarily attributed to the simultaneous ionization of the metastable and ground state, leading to an increased total ionization rate.

\section{\label{Sec:Summary}Summary}

In this study, both the Mo$^{15+}$/Mo$^{14+}$ ratio, the charge state distribution of Mo$^{14+}$ and Mo$^{15+}$ ions and visible spectrum of Mo$^{15+}$ ion were measured at the Shanghai low-energy EBIT, CUbic-eBIT (CUBIT). The energy level, radiative transitions probability, electron-impact excitation and electron-impact ionization cross sections of Mo$^{14+}$ and Mo$^{15+}$ ions were calculated using the Dirac-Fock-Slater method with a local central potential and distorted-wave approximation. All these calculation were based on the FAC package. Through a detailed analysis of various atomic processes, a theoretically most likely to occur channel was identified. A reasonable explanation was provided for the experimental observation of the Mo$^{15+}$ ion spectrum below the ionization energy of Mo$^{14+}$ ion (544.0 eV). The Mo$^{15+}$/Mo$^{14+}$ ions ratio was calculated using the collisional-radiative model, which considered the metastable state related process, with the resulting calculations showing strong agreement with the experiment.

\section{\label{Sec:ACKNOWLEDGMENTS}ACKNOWLEDGMENTS}
This work was supported by National Nature Science Foundation of China (Grant No. 12274352), the National Key Research and Development Program of China (Grant No. 2022YFA1602500).

\bibliography{apstemplate}
\end{document}